\documentstyle[aps,prl,psfig,twocolumn,amssymb]{revtex}
\begin{document}
\wideabs{
\title{Testing the Landau Fermi liquid description of the fractional
quantum Hall effect}
\author{Sudhansu S. Mandal and J.K. Jain}
\address{Department of Physics, 104 Davey Laboratory, The Pennsylvania
   State University, University Park, PA 16802}

\date{\today}

\maketitle

\begin{abstract}
We study theoretically 
the dispersion of a single quasiparticle or quasihole of the fractional quantum Hall
effect, obtained by injecting or removing a composite fermion.  By comparing to a 
free fermion system, we estimate the regime of validity of the 
Landau-Fermi-liquid-type description and deduce the Landau mass.
\end{abstract}
}

During the last decade, a Landau-Fermi-liquid type description of the fractional 
quantum Hall effect has been investigated in terms of composite fermions, namely 
electrons binding an even number of quantum
mechanical vortices \cite{book1,Jain,HLR,theory,Du,Shankar}.  
Numerous studies, both theoretical and experimental, 
indicate that this is a valid starting point\cite{book1}.  However, the 
extent as well as the regime of applicability of the Landau-Fermi liquid ideas 
is not entirely clear from a theoretical view point.  While for interacting electrons at zero
magnetic field the Landau-Fermi liquid theory is best justified asymptotically close to the Fermi
energy, for the compressible Fermi sea at
the half-filled Landau level, a logarithmic divergence of the mass has been predicted at very low
energies\cite{HLR,theory}, suggesting a breakdown of the Landau-type description in the limit of
vanishing temperature.  At the same time, it is also obvious that the description of the state 
in terms of composite fermions, and therefore also as a Landau Fermi liquid of composite fermions,
must become invalid at sufficiently 
high energies.  Therefore, one may ask what is the range of energies, if any, in which the
Landau-Fermi liquid type description may be valid for composite fermions.

\begin{figure}[t]
\vspace{-1.0cm}
\psfig{file=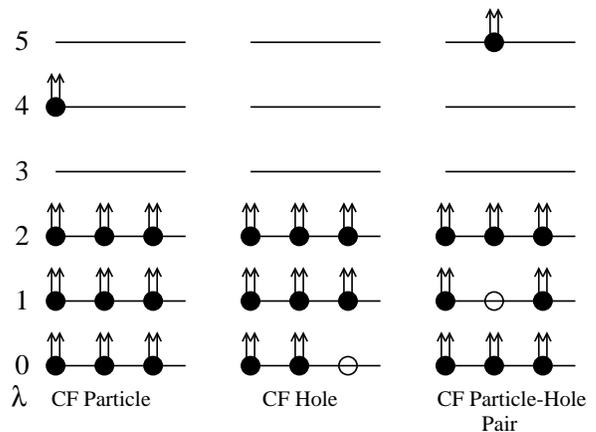,width=9.5cm,angle=-90}
\caption{Schematic depiction of a CF particle (in $\lambda=4$ CF-LL), a CF hole (in $\lambda=0$
CF-LL), and a CF particle-hole pair at
$\nu=3/7$.  The composite fermion is shown as an electron carrying two flux quanta.}
\label{fig1}
\end{figure}

The basic assumption of the Landau Fermi liquid theory is that the low energy excitation spectrum
of an interacting system resembles that of a system of free fermions.  
We consider here the fundamental building block, namely a single quasiparticle or quasihole 
of the FQHE state, the former
obtained by injecting a composite fermion into one of many unoccupied CF-LLs, and the latter by
removing a composite fermion from one of many occupied CF-LLs (see Fig.~\ref{fig1}).
We ask if the energy levels of the particle or hole can be consistently interpreted as the energy
levels of a single particle or hole in a free fermion system.
In Landau's Fermi liquid theory, the mass of the quasiparticle is defined by comparing its
dispersion to that of a quasiparticle in a free fermion system.  
By analogy, we interpret the energy level spacing of the composite fermion quasiparticle 
as an effective cyclotron energy to  
extract a mass for the composite fermion, which we call the Landau mass, $m^*_L$.

Our study shows that the analogy to a free fermion system  breaks down beyond a certain 
energy, which we crudely estimate to be $\approx 0.1 e^2/\epsilon l$, where
$\epsilon$ is the dielectric constant of the host material and $l=\sqrt{\hbar c/eB}$ is the
magnetic length.  We also find that the Landau masses of the particle and the hole are
in general different, but have substantial filling factor dependence and 
appear to approach the same value as the half filled Landau level is
approached.  (We note that because of the existence of a gap at the filling factors considered here,  
our work will not shed any light on the proposed logarithmic divergence at $\nu=1/2$.)  

Another mass for composite fermion has been defined in the past by interpreting the energy gap to
creating a particle hole pair of composite fermions as an effective cyclotron energy for the
composite fermion \cite{HLR,theory,Du,Shankar}.  In order to distinguish it from the Landau mass, 
it will be referred to it as the activation mass, $m^*_a$. The two masses are obviously distinct.
The Landau mass may not even be well defined, in general, but when it is, its value 
is deduced from the energy levels of a pre-existing particle or hole. 
The activation mass, on the other hand, is defined relative to a state containing no 
particle or hole, and thus also includes the self energies of the particle and the hole.
In Landau's theory, the constant self energy contribution 
is lumped together with the chemical potential and does not contribute to the mass.
The Landau and activation masses are relevant for different physical quantities.  The former, for
example, will be the mass that will appear in the specific heat.
The two masses are expected to be of similar magnitude, however.

\begin{figure}
\psfig{file=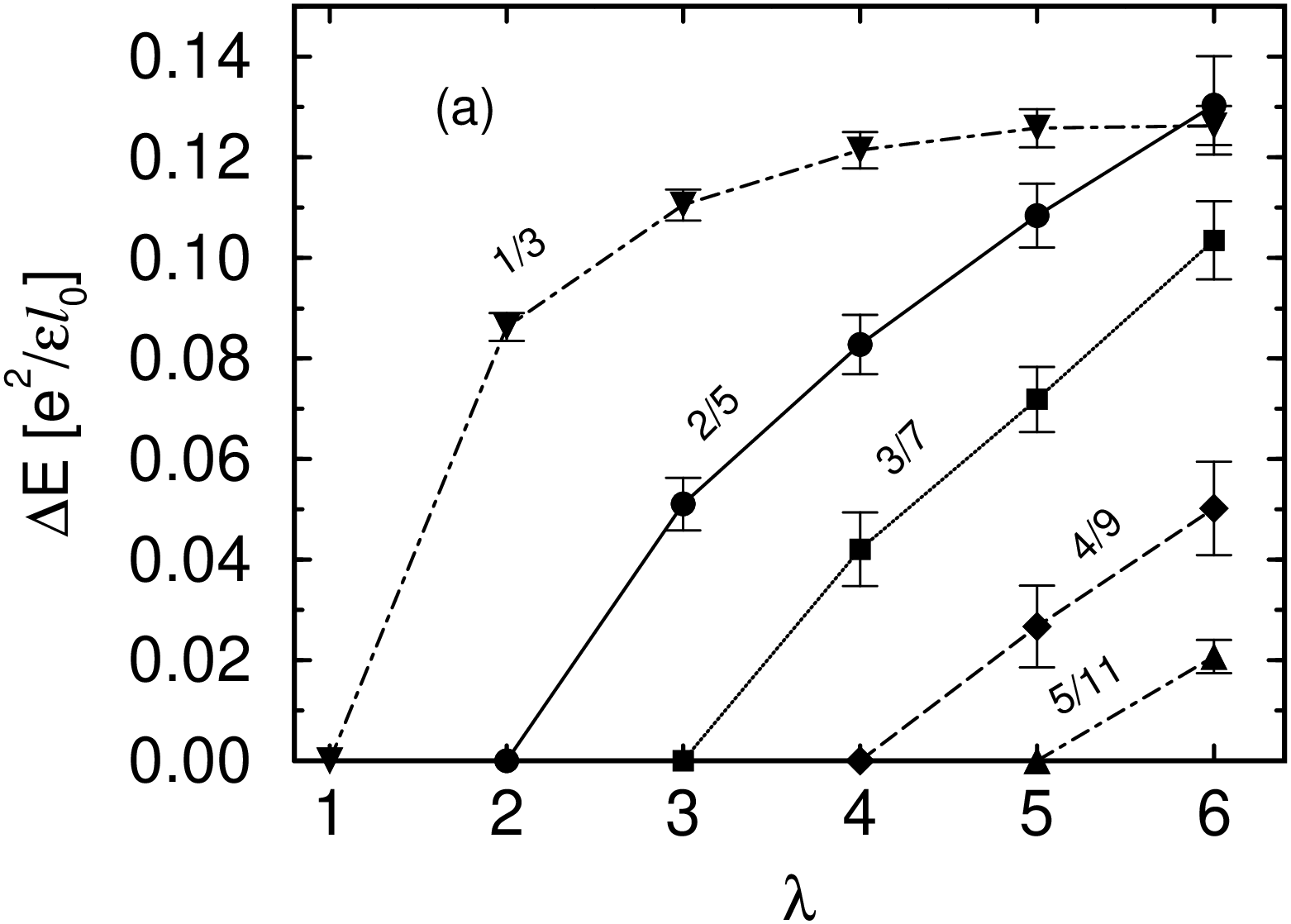,width=10cm,angle=-90}
\vspace{-1cm}
\psfig{file=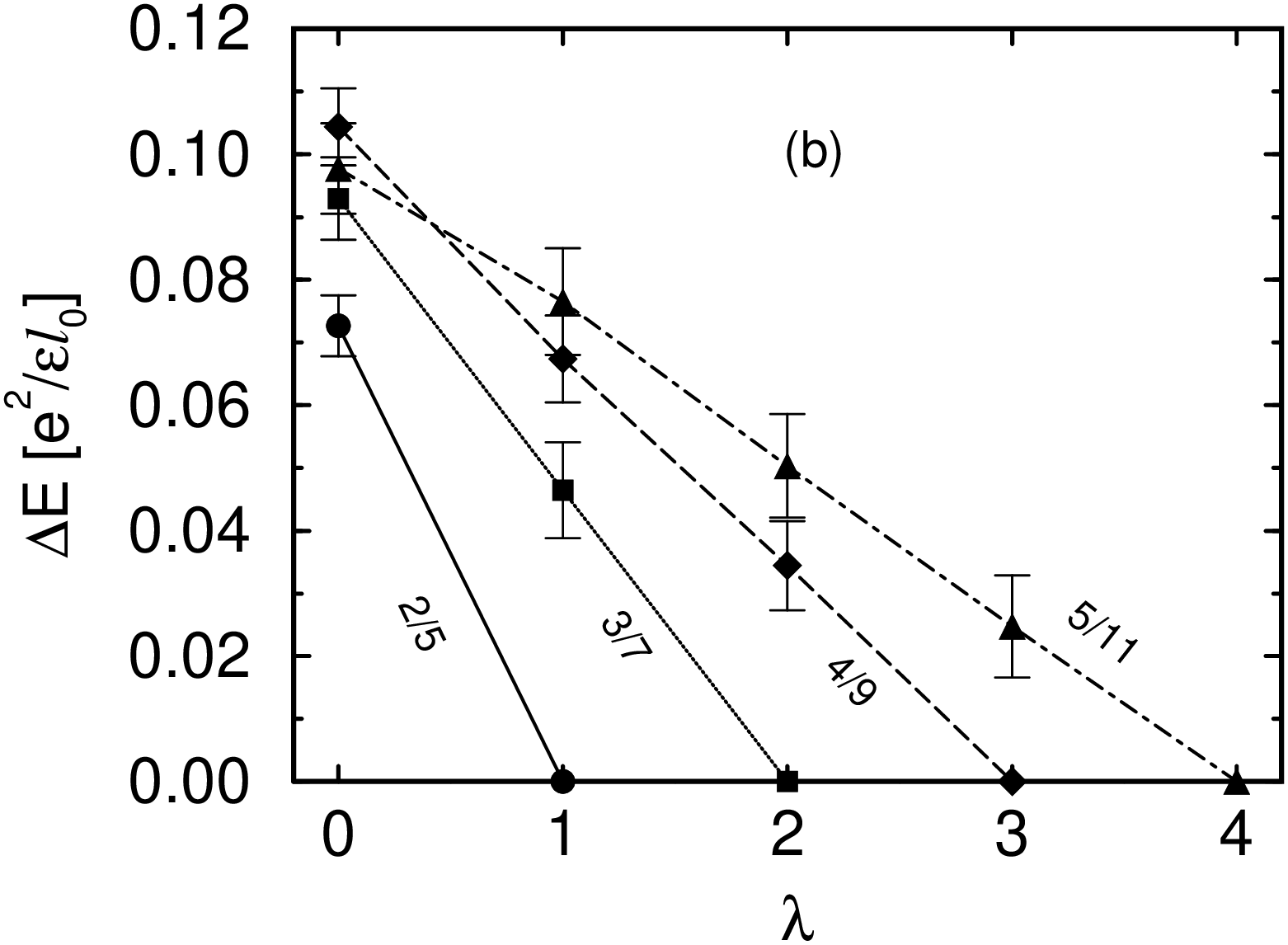,width=10cm,angle=-90}
\caption{The dependence of the 
energy of the CF particle (upper panel) and the CF hole (lower panel) on its Landau level
index, $\lambda$, at several filling factors. The energy is measured relative to the lowest
energy, and quoted in units of $e^2/\epsilon l_0$,
where $\epsilon$ is the dielectric constant of the background material, and $l_0=\sqrt{\hbar
c/eB}$ is the magnetic length. The error bar shows the uncertainty from Monte Carlo as well as
from the extrapolation to the thermodynamic limit.}
\label{fig2}
\end{figure}

Our calculations will be based on wave functions for the composite fermions,
which have been well tested in the past, and we expect the results to be quantitatively 
reliable for the FQHE states with strong gaps.  
The wave function for the ground state at $\nu=n/(2n+1)$ is given by 
$\Psi_{n/(2n+1)}=P_{LLL}\Phi_n \Phi_1^2$,
where $\Phi_n$ is the Slater determinant wave function for the state with $n$ filled Landau
levels, also denoted by $|n>$, and $P_{LLL}$ is the lowest LL projection operator.  
To construct a CF particle, we inject an additional
composite fermion in the $\lambda$th CF-LL, which corresponds to putting an additional electron in
the $\lambda$th LL at $\nu^*=n$.  The wave function of the CF particle is $
\Psi_{n/(2n+1),\lambda,m}^p=P_{LLL}
\Phi^{p}_{n,\lambda,m} \Phi_1^2$,  with $\Phi^{p}_{n,\lambda,m}=c^{\dagger}_{\lambda,m}|n>$,
where $c^{\dagger}_{\lambda,m}$ creates an electron in the $\lambda$th LL in the
state labeled by quantum number $m$.  (The Landau level index takes on values 
$\lambda=0,1,2,...$, and the state $|n>$ has all states occupied
with $\lambda\leq n-1$. Therefore, for a CF particle, we must have $\lambda\geq n$.) 
Similarly, the wave function for a CF-hole is given by $\Psi_{n/(2n+1),\lambda,m}^h=P_{LLL}
\Phi^{h}_{n,\lambda,m} \Phi_1^2$ where $\Phi^{p}_{n,\lambda,m}=c_{\lambda,m}|n>$, with $\lambda<n$.
We will work in the spherical geometry, in which $N$ electrons move on 
the surface of a sphere
under the influence of a radial magnetic field produced by a monopole of strength $q$, an integer
or a half integer, at the center.  The $\lambda$th Landau level is the shell with single particle
orbital angular momentum
$l=|q|+\lambda$, with single particle degeneracy $2l+1$.  For $N$ particles, the 
state $\Phi_n$ occurs at 
$ q_n = \frac{N-n^2}{2n}$, which corresponds to a total $q$ of $q=q_n+2 q_1$, with $q_1=(N-1)/2$,
for the ground state $\Psi_{n/(2n+1)}$.  This is a filled shell state, with total 
orbital angular momentum $L=0$.
The state $\Phi_n^p$ ($\Phi_n^h$) corresponds to  $q_n^p=\frac{N-1-n^2}{2n}$
($q_n^h=\frac{N+1-n^2}{2n}$), because now $N-1$ ($N+1$) particles completely 
fill $n$ Landau levels,  
giving the total monopole strength of $q^p=q_n^p+2q_1$ ($q^h=q_n^h+2q_1$) for the state 
$\Psi_{n/(2n+1),\lambda,m}^p$ ($\Psi_{n/(2n+1),\lambda,m}^h$).  It is important to note that 
the monopole strength is independent of the CF-LL index, $\lambda$, of the CF particle or hole.
The state $\Psi_{n/(2n+1),\lambda,m}^p$ ($\Psi_{n/(2n+1),\lambda,m}^h$) has total angular momentum
%When a CF particle or a CF hole is added to the $\lambda$th CF-LL, the angular momentum of the
%state is 
$L=q^p+\lambda$ ($L=q^h+\lambda$), $m$ denoting the z component of $L$.  Since the
energy is independent of $m$, we work with the state $m=L$.  An advantage of the
spherical geometry is that the states with CF-particle or CF-hole in different CF-LLs (i.e.,
with different $\lambda$) are automatically orthogonal 
on account of their different orbital symmetry.  The various Slater determinants $\Phi_n$ 
can be readily
constructed with the knowledge of the single particle eigenstates.
Their explicit forms of the lowest LL projected wave functions has been given previously \cite{JK}.

The energies of the CF particle and the CF hole can be expressed as 
multidimensional integrals
that are evaluated by Monte Carlo requiring up to $10^7$ steps.
Particle and hole occupying CF Landau levels up to $\lambda\leq 6$
are considered.  In each
case, the thermodynamic limit has been extracted by studying systems with up to 46 particles.
The results for up to $\nu=5/11$ are shown in Fig.~(\ref{fig2}). 
The analogous plot of the energy levels of a free fermion system will have straight lines (with
slope equal to the cyclotron energy), so a deviation from a straight line indicates a breakdown  
of the free fermion desctiption.  At $\nu=1/3$, the cyclotron
gap changes even for the lowest energy levels, which implies that  
the Landau Fermi liquid description is not valid here and no mass can be defined.
It would be troublesome if the energy levels continued to behave similarly at other fractions.
However, for other fractions, especially for $n\geq 3$, the cyclotron
energy becomes reasonably constant close to the Fermi energy.
Taken all together, the results suggest that the free fermion 
description of the system is valid up to an energy $E\sim 0.1 e^2/\epsilon l$, but breaks
down beyond that.  This conclusion obviously extends to states with a sufficiently dilute 
concentration of particles and holes, when the interaction between them is negligible.

\begin{minipage}{75mm}
\begin{table}[t]
\caption{The effective cyclotron energy, $\hbar\omega_c^*$, (in units of $e^2/\epsilon l_0$) 
for the CF particle and the CF hole at several filling factors.  Also given is the normalized
Landau mass\protect\cite{footnote2}, defined by the equation $m^*_{L,nor}=m_L^*/ (m_e \sqrt{B[T]})$. 
\label{tab1}}
\begin{center}
\begin{tabular}{|c|c|c|c|c|} {\bf $ \nu$ } &   \multicolumn{2}{c|}{ \bf CF particle} & 
\multicolumn{2}{c|}{\bf CF hole} \\  
\cline{1-5}
 &    $\hbar\omega_c^*$ & $m^*_{L,nor}$ & $\hbar\omega_c^*$  & $m^*_{L,nor}$ 
\\ \hline
$2/5$  & - &  - &  0.0727(48) & 0.073(5)  \\ \hline
$3/7$  & 0.0356(40) & 0.106(12) & 0.0465(30) & 0.081(5) \\ \hline
$4/9$  & 0.0255(40) & 0.115(18) & 0.0351(47) & 0.084(11) \\ \hline
$5/11$ & 0.0208(33) & 0.115(18) & 0.0248(37) & 0.097(14) \\
\end{tabular}
\end{center}
\end{table}
\end{minipage}

We determine the particle and hole masses from the slope of the points below $E\sim 0.1
e^2/\epsilon l$ for several fractions.  The 
results are given in Table~\ref{tab1}. The masses are different for the particle and the hole.
They are in general also filling factor dependent, but  
seem to be approaching the same value 
in the limit of $\nu=1/2$, as expected.  A simple extrapolation produces a CF mass of 
$m^*_{nor}\approx 0.11-0.13$ at the half-filled Landau level.  Such an 
extrapolation must obviously be treated with the understanding that it does not capture  
a logarithmic divergence of the mass at $\nu=1/2$ predicted in other approaches \cite{HLR} or a
much stronger divergence indicated by experiment.  The extrepolation in the present work 
may give the mass at an intermediate range of temperaturs, though.
The theoretical value for the activation mass (for Coulomb interaction) for various FQH states
is smaller but of similar magnitude ($m^*_{a,nor}\approx 0.08$\cite{HLR,JK}).

The existence of a well defined Landau mass is non-trivial for composite fermions.
The mass of the quasiparticle in ordinary Fermi liquids is subject to perturbative 
corrections due to interparticle interactions.  In contrast, even though it is natural to attribute
a mass to composite fermions, given that we speak of their Fermi seas and Landau levels, 
the situation is subtle here because the mass is a non-perturbative effect:  the original 
problem, namely interacting electrons in the lowest
Landau level, has no mass parameter, and the mass of the composite fermion
is generated entirely from the inter-electron interactions.
There is no simple, solvable limit in which the mass of the composite fermion is known.

Does the above insight into the energy levels of a single particle or hole relate in any
way to the structure of the energy spectrum of  
more complicated excitations, for example a particle hole excitation (PHE) of the incompressible 
state at $\nu=n/(2n+1)$?  Motivated by this question, we 
proceed to study the PHE spectrum, also known as the 
``single particle excitation spectrum," of 
composite fermions, which is the allowed range of energy for the excitation of a single 
composite fermion out of the ground state, leaving behind a hole.
The wave function for such a state is given by $P_{LLL}
\Phi^{p-h}_{n,\lambda,m,\lambda',m'} \Phi_1^2$ where $\Phi^{p-h}_{n,\lambda,m,\lambda',m'}=
c^{\dagger}_{\lambda,m}c_{\lambda',m'}|n>$.  The situation is now significantly 
more complicated for two reasons.  First,  these states do not automatically have the correct
symmetry, because they are not the eigenstates of $L$ and $L_z$.  This can be taken care of 
straightforwardly by working instead with appropriate linear combinations, 
$\sum_{m,m'}<LM|q_n+\lambda,m; q_n+\lambda',m'>
\Phi^{p-h}_{n,\lambda,m,\lambda',m'}$.
Now, we must diagonalize the Coulomb Hamiltonian in the
Hilbert space defined by these states. The number of states at a given $L$ is finite because 
the minimum $L$ of a particle hole pair is $\Delta \lambda=|\lambda-\lambda'|$.  
However, 
these states are in general not orthogonal.  We therefore first obtain an orthogonal basis
according to the Gram-Schmid procedure.  For this purpose, we need to evaluate
the off-diagonal projections $\big< u_r \vert u_s \big>$, where $\{\vert u_r \big>\}$ denotes
the set of non-orthogonal basis states.
Since the Monte Carlo method is most efficient when the integrand is positive definite,
we determine $\big< u_r \vert u_s \big>$ ($r\neq s$) by using the relation  
\cite{Bonesteel}:
$$\big< u_r \vert u_s \big> = \frac{ \big< u_r + u_s \vert u_r +u_s \big>
      - \big< u_r \vert u_r \big> - \big< u_s \vert u_s \big> }{2}.$$

\begin{figure}
\psfig{file=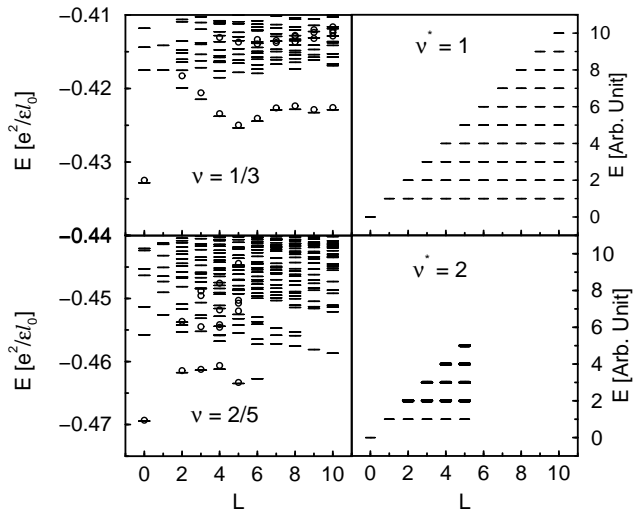,width=9cm,angle=-90}
\caption{The particle hole excitation spectrum of composite fermions (circles, left panels) 
for $N=10$ particles at $\nu=1/3$ and 2/5.  The dashes are exact eigenenergies, taken from
Ref.~\protect\onlinecite{He}.  The right panels
show the single particle excitation spectrum at $\nu^*=1$ and $\nu^*=2$ for non-interacting
fermions, in units of the cyclotron energy.  The darker dashes in the 
lower right hand panel denote two degenerate multiplets, 
corresponding to excitations out of the two occupied CF Landau levels.} 
\label{fig3}
\end{figure}

\noindent
The off-diagonal matrix elements of the Hamiltonian are evalulated analogously, a diagonalization
of which produces the energy eigenvalues in the restricted basis of states containing a single CF
particle-hole pair.  The results are in principle exact in the subspace of states containing
only a single particle hole pair; the statistical uncertainty is estimated to be 
smaller than the symbols used in Fig.~(\ref{fig3}).

\begin{figure}
\vspace{-2.5cm}
\psfig{file=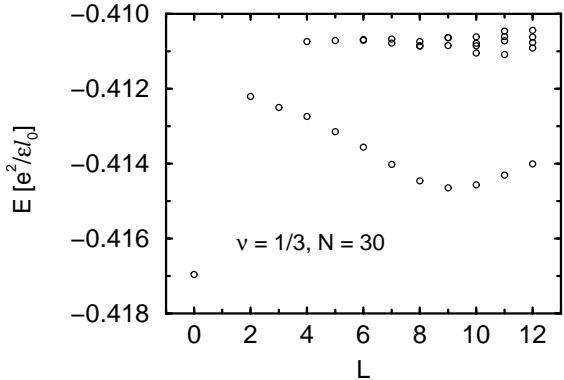,width=11cm,angle=-90}
\caption{The single particle excitation spectrum for composite fermions at $\nu=1/3$ for $N=30$
particles.}
\label{fig4}
\end{figure}

Fig.~(\ref{fig3}) shows the PHE spectrum of composite fermions at $\nu=1/3$ and $\nu=2/5$ for
$N=10$ particles, along with the
PHE spectrum at $\nu^*=1$ and $\nu^*=2$, which is what one would expect in a model of
non-interacting composite fermions with a well defined mass.
Fig.~(\ref{fig4}) shows the PHE spectrum at $\nu=1/3$ for a much bigger system \cite{footnote}.
Again, the basic message is that the analogy to free fermions breaks down above $\sim 0.1 e^2/\epsilon
l_0$.  Certain qualitative features of the PHE spectrum, which might 
appear puzzling at first sight, 
can be understood in light of our earlier results.  For example, at $\nu=1/3$, the first
PHE branch is well defined, but the higher branches merge into one another
(Figs.~\ref{fig3} and \ref{fig4}), which can be understood as a consequence of the fact that
at $\nu=1/3$, the effective cyclotron energy of composite fermions decreases rapidly as the CF-LL
index increases.  At $\nu=2/5$, the effective cyclotron energy is better defined
(Fig.~\ref{fig3}), consistent with a more spread out PHE spectrum.

As expected from previous work \cite{Wu2}, there are fewer states in the actual PHE spectrum of
composite fermions than what one would expect from the naive model.   The reason is that many
states at $\nu^*=n$ are annihilated upon multiplication by $\Phi_1^2$ followed by lowest Landau
level projection.  The naive independent CF model thus contains spurious states that have no
counterpart in the exact spectrum.  The number of spurious states in the independent CF model 
increases with energy, again indicating that the model may be valid only for low energy
states. The effective independent fermion description appears to be valid only for the 
low energy solutions, namely the lowest band, and also the first excited band (providing one
neglects the thermodynamically insignificant spurious state at $L=1$). 

This work was supported in part by the National Science Foundation under grant no.
DMR-9986806.  We thank the Numerically Intensive Computing Group led by V.
Agarwala, J. Holmes, and J. Nucciarone, at the Penn State University CAC for
assistance and computing time with the LION-X cluster,
and G. Murthy and R. Shankar for useful discussions.


\begin{references}
\bibitem{footnote2}
At electron filling $\nu=n/(2n+1)$, the system maps into composite fermions
at an effective filling $\nu^*=n$, with an effective magnetic field $B^*=B/(2n+1)$.  
The Landau mass, $m_L^*$, is defined by $\omega_c^*\equiv eB^*/m_L^* c$.
Assuming that electrons are confined to their lowest
LL, the only energy scale in the problem is the interaction energy, and therefore we must have 
$\hbar\omega^*_c=\delta (e^2/\epsilon l_0)$, where 
$l_0\equiv \sqrt{\hbar c/eB}$ is the magnetic length,
$\epsilon$ is the dielectric constant of the background material (which will be taken to be
$\epsilon=12.8$, as appropriate for GaAs), and $\delta$ is the dimensionless cyclotron energy
in units of
$e^2/\epsilon l_0$.  For GaAs, the relation $m_L^*=\hbar^2 \epsilon /[(2n+1)\delta e^2 l_0]$ 
can be conveniently expressed as ${m_L^*}/{m_e}\equiv m^*_{L,nor}\sqrt{B[T]},$
where $m_e$ is the electron mass in vacuum, $B[T]$ is the magnetic field measured in Tesla, and
the ``normalized" Landau mass is given by $m^*_{L,nor}=0.0264/\{(2n+1)\delta\}$ for GaAs.
Here, $m_e$ is only a convenient unit; the composite fermion mass is not
related to it in any way.
\bibitem{footnote}  We have considered only the CF LLs with $\lambda\leq 6$.  
For $\nu=2/5$, this restricts us to particle-hole pairs with $L\leq 5$.  However, 
for $\nu=1/3$, we are able to go to much larger $L$, thanks to the fact that the states 
involving excitations into high CF LL's are spurious, as explained in the text.
\end{references}
\end{document}